\documentstyle[12pt]{article}
\newcommand{\beq}{\begin{equation}}
\newcommand{\eeq}{\end{equation}}
\newcommand{\e}{\varepsilon}

\newcommand{\bpsi}{{\bar\psi}}

\newcommand{\la}[1]{\label{#1}}

\newcommand{\th}{\mbox{'t~Hooft }}
\begin{document}
\begin{flushright}
{\bf RUB-TPII-17/94}
\end{flushright}
\begin{center} {\large \bf
${\bar p} p$ ANNIHILATION: MANIFESTATIONS OF
INSTANTON INDUCED QUARK PAIR PRODUCTION ? }\\[0.25cm]
\vspace{0.7cm}
 M.~Polyakov \footnote{Permanent address:
{\it  Theory Division of Petersburg Nuclear Physics Institute,
188350 Gatchina, Russia}} and
 M.~Shmatikov \footnote{Permanent address:
{\it
 Russian Research Center "Kurchatov Institute", 123182 Moscow, Russia}}
\\
\vspace{0.75cm}
{\it Institut f\"ur Theor. Physik II, Ruhr-Universit\"at Bochum,
D-44780 Bochum, Germany} \\

\end{center}
\begin{abstract}
Production of strangeness in the low-energy  ${\bar p} p$ annihilation
is analyzed as a manifestation of the instanton induced multiquark
vertex. Approach based on the {\it qualitative} properties of the
\th mechanism provides a unified description of available experimental
data on $K$- and $\phi$-meson production. Crucial tests for the hypothesized
mechanism are suggested.
\end{abstract}

\section{Introduction}
Success of the naive quark model in describing properties of
the nucleon prompts the conclusion that the nucleon wave function
is dominated by the configuration containing three
(constituent) quarks. The latter is generally believed
to be an object emerging from the current quark due to the spontaneous
chiral symmetry breaking. However, success of the constituent
quark model does not provide an insight in the (nonperturbative) QCD
mechanisms, the sea quarks and gluons being 'hidden' in the
constituent quark. It is natural to assume that a clue to
understanding nonperturbative phenomena inherent to QCD may
be provided by mechanisms driving production of (constituent) quark-antiquark
pairs. Not all of the plethora of
conceivable processes are equally suitable to our purposes.
Crudity of the state-of-art understanding of QCD mechanisms
does not allow to get reliable quantitative results. This makes us
to choose reactions where channels with hadron production
are exhibited in their most clear-cut form. The best candidate
seems to be the ${\bar p}p$ annihilation where the limited
understanding of strong-interaction mechanisms operative at low energies
presumably is not  distorted by high-energy effects.

For further consideration it is convenient to distinguish
perturbative and nonperturbative mechanisms of (constituent) quark-pair
production. Well known example of a nonperturbative
mechanism is the instanton-induced effective quark interaction
materialized as the celebrated 't Hooft vertex \cite{tHooft}.
A purpose of the present research is to draw qualitative conclusions
concerning existence of this mechanism for
observables in the ${\bar p}p$ annihilation.
\section{'t Hooft interaction}
One of the most prominent features of  strong interactions is the
spontaneous
breaking of the chiral symmetry. This phenomenon occurs due to non--trivial
structure  of the QCD vacuum which is populated by topologically
non-trivial gluon fluctuations -- (anti)instantons.
The averaging over instanton ensemble induces many-quark interactions
first suggested by 't Hooft\cite{tHooft}.
 We quote here the result for the  $N_f=3$ case which seems to be of prime
interest for the low-energy phenomena.
In this case one has a 6-fermion vertex of the following structure:
\[
{\cal M } \sim \e^{f_1f_2f_3}\e_{g_1g_2g_3}
\times
\left[
\left(\bpsi_{Rf_1}\psi_L^{g_1}\right)
\left(\bpsi_{Rf_2}\psi_L^{g_2}\right)
\left(\bpsi_{Rf_3}\psi_L^{g_3}\right)
\right. +
\]
\beq
\left.
+\frac{3}{8(N_c+2)}
\left(\bpsi_{Rf_1}\psi_L^{g_1}\right)
\left(\bpsi_{Rf_2}\sigma_{\mu\nu}\psi_L^{g_2}\right)
\left(\bpsi_{Rf_3}\sigma_{\mu\nu}\psi_L^{g_3}\right)
\right],
\la{vertex}
\eeq
where $f(g)=1\ldots N_f$ stands for flavor index.

Inspection of the above expression reveals immediately the most important
{\it qualitative} properties of this 6--quark vertex which will be
exploited in the analysis below:

\begin{itemize}
\item
it contains quarks of all flavors ($u$, $d$ and $s$) and as such
incorporates intrinsically strangeness production
\item
${\bar q}q$ pair of each flavor enters the vertex one and only one time
\end{itemize}
These qualitative features have far-reaching consequences. Indeed,
let us for clarity adhere to the naive quark model assuming that
nonstrange hadrons are built from $u$- and $d$-quarks only (admixture,
if any,
of the strange sea quarks in the wave function of, say, the nucleon is
anyway small and does not affect our conclusions). Then
whenever interaction of nonstrange hadrons involves quark-pair
production the 't Hooft mechanism ensures that the final state
contains ${\bar s}s$ pair. Stated differently the vertex (\ref{vertex})
operating in the coupling of nonstrange hadrons produces either
a pair of strange mesons or a meson with the hidden strangeness, i.e.
the $\phi$-meson. Thus, investigation of strange particle production
proves to be a selective probe for nonperturbative mechanisms of
strong interaction. These qualitative conclusions are naturally
obscured by presence of other dynamic mechanisms which are 'flavor blind'
and as such do not distinguish strange quarks. In absence of
strong-interaction theory  in the nonperturbative domain an attempt
to separate the contribution of the 't Hooft mechanism would be
hopeless weren't it for its multi-quark nature. Translated into
the language of quark-pair production it may be considered as a
mechanism changing the number of ${\bar q}q$ pairs by unity.

To assess importance of the 't Hooft mechanism and its (relative)
contribution let us consider in more detail what can be learned
from the success of the constituent quark model. The basic assumption of the
model is that low-lying hadrons have well defined quark structure,
nucleons and mesons being $3q$ and ${\bar q}q$ states respectively.
Description of the nucleon
in the low $Q^2$ domain
as a conglomerate of constituent $u$- and $d$-quark implies
that the admixture of explicit quark-antiquark pair in its
wave function is not large and can be neglected for our qualitative
analysis. This conclusion can be rephrased as a statement that
quark-production mechanism plays a minor role in coupling of constituent
quarks within the nucleon. Thus we are
guided to look for 't~Hooft mechanism manifestations in processes
involving mandatory change of ${\bar q}q$ pairs number wherein
it hopefully can be competitive with other mechanisms (hereafter for
brevity we shall call them 'conventional' ones).
To be more specific we focus attention on processes with
nonstrange particles in the initial state.
Within this (maybe over)simplified picture we can conclude that
the 't~Hooft mechanism is expected to be distinguished in
processes involving 6 quarks (or antiquarks in various combinations -
see below) whenever the number of quark pairs changes by unity.
Processes satisfying these 'counting rules' are marked by noticeable
admixture of ${\bar s}s$ pair in the final state.

\section{Strangeness production in \mbox{${\bar p} p$} annihilation}
It was argued in the Introduction that
${\bar p}p$ annihilation seems to be the process most suitable for
probing instanton induced mechanism, since it allows production of
mesons  within
the low-energy region and, moreover, it provides a variety of final
states which can be
subject to the analysis. To avoid problems related to complicated
dynamics of quarks merging into hadrons we analyze the ratios of
matrix elements for the processes with and without strange quarks in
the final state. More specifically, we consider the ratio
\beq
\rho = \frac{\vert M({\bar p}p\rightarrow \{{\bar s}s\} + hadrons)\vert}
{\vert M({\bar p}p\rightarrow\{{\bar q}q\} + hadrons)\vert}
\label{not}
\eeq
Here the numerator stands for the amplitude of the process of the
${\bar p} p$ annihilation into
the state containing a pair of strange quarks (denoted symbolically
$\{{\bar s}s\}$) either in the form of strange hadrons or mesons
with hidden strangeness virtually accompanied by nonstrange
hadrons. Denominator of (\ref{not}) denotes the
amplitude of the similar process where strange-quark pair is substituted
by light quarks in the state with the same quantum numbers. The
value of $\rho$ ratio will presumably be less sensitive to the
dynamics of quark merging into hadrons which is irrelevant for
the problem under consideration.

There is the basic difference between processes of strange quark-production
and ${\bar p} p$ transition into the nonstrange state which are utilized
 as benchmarks for the former. Presence of strange quarks in the
final state unambiguously signals that annihilation and
production mechanisms are operative. However, this is not the case
when the final state contains light
quarks only. Recall that according to the
constituent quark model the nucleon consists of three quarks while
a meson is a bound ${\bar q} q$ state. With this bookkeeping in mind
we can divide ${\bar p} p\rightarrow nonstrange \:\: hadrons$
\footnote{Unless specifically stated we do not make difference
between channels containing pair of strange particles and a
meson with hidden strangeness} processes into
two classes. First of them comprises channels where the number of
quark-antiquark pairs in the final state differs from that of in the
initial configuration ($\Delta n\neq 0$). These 'truly annihilation'
processes necessarily
involve mechanisms of quark-pair annihilation and production. The
other class involves processes with coinciding number of ${\bar q} q$
pairs in the initial and final states ($\Delta n = 0$). They
can be contributed by various mechanisms: those involving quark pair
annihilation
and non-annihilation mechanisms. Rearrangement and scattering may be
considered as representative examples of the latter. Successful
description of low-lying hadron properties in the framework of
the constituent quark model prompts us to conclude that whenever both
mechanisms are present the dynamics of the system is dominated by
non-annihilation ones, implying that the number of quark-antiquark pairs
is  not  changed.

Two types of ${\bar p} p$ interaction dynamics map onto two different
scales of the $\rho$ ratio defined in (\ref{not}). When
formulated in terms of quark coupling, it exhibits the relative strength of
strange-quark production either to  the production of light-quark pair or
(depending on the type of the benchmark process -- see above), to the
strength of a non-annihilation mechanism. Addressing experimental data
on ${\bar p} p$ annihilation into  final states which belong to
different classification classes we fix scales of the $\rho$ ratio
(\ref{not}) to be used afterwards for the analysis of other processes.
Let us begin with the \mbox{${\bar p} p\rightarrow {\bar \Lambda}\Lambda$}
process. Its nonstrange counterpart \mbox{${\bar p} p\rightarrow {\bar p} p$}
belongs to the class of processes dominated by non-annihilation
mechanisms. Then it would be natural to expect that the corresponding
value $\rho_{0}$ is small (the subscript is a mnemonics indicating the
minimal change of the number of quark-antiquark pairs in the nonstrange
channel). Using experimental data on cross sections of
both processes (\mbox{$\sigma_{tot}({\bar p}p\rightarrow
{\bar\Lambda}\Lambda)\approx 120 \mu$b}, see e.g. \cite{Barnes} and
\mbox{$\sigma_{el}({\bar p}p)\approx 50$ mb} \cite{Who}) we, indeed,
get a small value
\beq
\rho_{0} = \frac{\vert M({\bar p}p\rightarrow {\bar\Lambda}\Lambda)\vert}
{\vert M({\bar p}p\rightarrow {\bar p}p)\vert}\approx 0.05\, .
\label{ll}
\eeq
Representative example of processes where annihilation plays, in contrast,
noticeable
role is furnished by the ${\bar p}p$ annihilation into $K$- and $\pi$-meson
pairs. In this case both channels involve production of a quark-antiquark
pair.  This is a
signal that the corresponding $\rho_{1}$ ratio may be rather
large. Invoking experimental data \cite{Koenig} on the corresponding
branching ratios (with phase-space corrections included) we get
\beq
\rho_{1}= \frac{\vert M({\bar p}p\rightarrow K^+K^-)\vert}
{\vert M({\bar p}p\rightarrow \pi^+\pi^-)\vert} \approx 0.3\div 0.5
\label{br}
\eeq
Note the significant difference of $\rho_0$ and $\rho_1$ values which
reflects the difference of the dynamics underlying two  types
of annihilation process.
This observation
provides a clue to understanding a challenging problem of
the $\phi$-meson (hidden strangeness) production in the
${\bar p}p$ annihilation and the role played by the 't~Hooft mechanism.

The problem under consideration is related to a surprisingly high yield
of $\phi$-mesons
(in the ${\bar p}p\rightarrow\phi\pi$ process) as compared to the similar
reaction
with $\phi$-meson substituted by the $\omega$-meson \cite{Asterix}.
'Abundance' of $\phi$-mesons is explained either by assuming sizeable
admixture of strange sea in the nucleon wave function \cite{Ellis},
\cite{Decker}
(conclusions of the former are criticized, however, in \cite{Storrow})
 or by means of  'conservative' mechanism involving production
and rescattering of $K$-mesons \mbox{${\bar p}p\rightarrow
K^*\bar{K}\rightarrow\phi\pi$} \cite{Lev}. However, the most vulnerable
point of these mechanisms is that they
fail to reproduce simultaneously relatively small branching ratio
of the $\phi$-meson production when accompanied by two $\pi$-mesons
(see \cite{Koenig}). Analysis of reaction mechanisms given above
shows that two of these mechanisms belong to different classes.
Inspection of quark content of the $\phi\pi$ and $\omega\pi$ final states
shows
that they belong to the annihilation ($\Delta n = 1$) class. Then
we conclude that the ratio of corresponding amplitudes is about the
same as the ratio of \mbox{${\bar p}p\rightarrow K^+K^-$} and
\mbox{${\bar p}p\rightarrow \pi^+\pi^-$}
annihilation rates (see (\ref{br}))
\beq
\frac{\vert M({\bar p}p\rightarrow \phi \pi)\vert}
{\vert M({\bar p}p\rightarrow \omega\pi)\vert} \approx
\rho_{1}\approx 0.3\div 0.5
\label{lll}
\eeq
Obtained ratio of matrix elements corresponding to the branching ratio
\mbox{$10\div 20\%$} agrees well with the experimental observations
\cite{Koenig},
\cite{Asterix}. The \mbox{${\bar p}p\rightarrow\omega\pi\pi$} process
belongs in contrast to the non-annihilation type ($\Delta n = 0$). Hence
the ratio of matrix elements reproduces the pattern of the eq.(\ref{ll})
yielding
\beq
\frac{\vert M({\bar
p}p\rightarrow \phi \pi\pi)\vert} {\vert M({\bar p}p\rightarrow
\omega \pi\pi)\vert }\approx\rho_0
\approx 0.05.
\label{brr}
\eeq
Corresponding branching ratio is about $0.2\div 0.3\%$. This value is
somewhat lower than the
experimental one $\approx 0.7\%$ \cite{Asterix}, however, the
qualitative features of experimental data are reproduced rather well.

\section{Manifestations of the \th  mechanism }
Up to now we have not specified yet the role played by instanton
induced annihilation mechanism. Recalling properties of the \th
 vertex
discussed in the Section~2 is natural to make an assumption
that it is a driving mechanism of the strangeness production.
This hypothesis complies well with the observations made in the
previous section. Thus the instanton induced interaction provides
a natural explanation of the experimental data. However, identifying
the \th vertex with the dominant source of strangeness production
enables us to make in addition predictions which are based on the {\it
qualitative} features of this vertex. Analysis of ${\bar p}p$ interaction
indicates that  more easily observable manifestations of the
(strange-quark producing) \th vertex are
to be expected in the 'truly annihilation' processes
($\Delta n = 1$). To separate contribution coming from the 't~Hooft
mechanism let us recall its peculiar properties. First, the
't~Hooft vertex involves the ${\bar s}s$ pair but, in contrast
to light quarks, it enters one time only (see (\ref{vertex})).
It implies that the
instanton induced mechanism under consideration cannot produce
a final state containing two ${\bar s}s$ pairs. At the
same time conventional mechanisms of quark-pair production,
being 'flavor blind' do not involve a suppression related to strangeness.
Thus, a crucial test for instanton induced vertex
is given by the processes containing either four $K$-mesons or
two $\phi$-mesons in the final state. If bulk of the final-state
strangeness is provided by the 't~Hooft mechanism the amplitude
of the $\phi\phi$ production is strongly suppressed
\beq
\frac{\vert M({\bar
p}p\rightarrow \phi\phi)\vert} {\vert M({\bar p}p\rightarrow
\omega\omega)\vert }\ll 1
\label{ffffk}
\eeq
If, in contrast, ${\bar q}q$ pairs are produced predominantly by a
'flavor-blind' mechanism the counterpart of the $(\ref{ffffk})$ will
apparently correspond to the pattern of $K$- and $\pi$-meson
production reading (see (\ref{br}))
\beq
\frac{\vert M({\bar
p}p\rightarrow \phi\phi)\vert} {\vert M({\bar p}p\rightarrow
\omega\omega)\vert }\approx\rho_1\approx 0.3\div0.5
\label{fka}
\eeq
Available experimental data indicate that the first estimate
(\ref{ffffk}) holds true. Indeed, the branching ratio of the
${\bar p}p\rightarrow \omega\omega$ annihilation is about
$3.3\cdot 10^{-2}$ \cite{brom}. The total cross section of
${\bar p}p$ interaction in the low-energy region (up to
$\sqrt s \approx 2$~GeV) is about $100\div200$~mbarn, yielding
cross section of the $\omega\omega$ production
\mbox{$\sigma ({\bar p}p\rightarrow \omega\omega)\approx 3\div6$~mbarn}.
This value is to be confronted to the cross section of the
$\phi\phi$ pair production at the same energy which is equal to
\mbox{$\sigma ({\bar p}p\rightarrow \phi\phi)\approx
 (1.5\div 4) \cdot10^{-3}$~mbarn}
\cite{phiphi}.
The ratio of matrix elements corresponding to values of cross
section under consideration equals
\beq
\frac{\vert M({\bar
p}p\rightarrow \phi\phi)\vert} {\vert M({\bar p}p\rightarrow
\omega\omega)\vert }\approx 0.01\div0.04
\label{fkb}
\eeq
Obtained ratio strongly disagrees with the value of
$\rho_1\approx 0.3\div0.5$ as predicted by ($\ref{fka}$) when
strange quark pairs are \underline{not} produced by the \th
mechanism. This result may be considered as a strong indication
that strangeness production is indeed dominated by the instanton
induced vertex.

Predictions similar to  (\ref{ffffk}) and (\ref{fka})
can be made as well for the
\mbox{${\bar p}p\rightarrow {\bar K} K{\bar K} K$} amplitude as
compared to the
amplitude of \mbox{${\bar p}p\rightarrow\pi\pi\pi\pi$} production.

Another specific property of the 't~Hooft vertex is that it can change
the number of quark-antiquark pairs not more than by unity. This property
enables us to conclude that the presence of the instanton induced
mechanism can be signalled by the value of the magnitude of strangeness
production accompanied by at least 4 pions. Indeed, consider the
\mbox{${\bar p}p\rightarrow \phi+ 4\pi$} annihilation channel.
Assuming that quark-pair production is driven by conventional mechanisms
we get a conservative estimate (compare to (\ref{lll})):
\beq
\frac{\vert M({\bar p}p\rightarrow \phi+ 4\pi)\vert}
{\vert M({\bar p}p\rightarrow \omega+ 4\pi)\vert}\approx
\frac{\vert M({\bar p}p\rightarrow \phi+\pi)\vert}
{\vert M({\bar p}p\rightarrow \omega+\pi)\vert}\approx 0.3\div 0.5
\label{fp}
\eeq
If, alternatively, quark-pair production is driven by the instanton
induced vertex the ratio of amplitudes in (\ref{fp}) will be strongly
suppressed.

Resuming, we have suggested a solution to challenging problems related
with strangeness production in the ${\bar p}p$ annihilation. Our approach
bases on the assumption that (strange) quark-pair production is controlled by
the instanton induced mechanism. Peculiar properties of the 't~Hooft
vertex provide an elegant explanation of experimentally observed
regularities and, moreover, enable determination of decisive tests.
It should be stressed that the developed approach \underline{does not}
require an admixture of the strange-quark sea in the nucleon wave
function.
Annihilation processes with the strangeness production is customary
to analyze in terms of the celebrated OZI rule. It should be
stressed that this phenomenological rule is satisfied in the
vector and tensor channels,
whereas in the scalar and pseudoscalar channels it is known to be
strongly violated. Observed violations of the OZI rule
can be attributed  to the contribution of so called
'direct instantons' \cite{inst}.
These results indicate unambiguously
that spin degrees of freedom are deeply involved in the strangeness
production mechanisms. The multifermion \th vertex treated in our
approach as the dominant source of ${\bar s}s$ pairs possesses specific
spin properties: inspection of (\ref{vertex})) shows immediately
that quarks and antiquarks (of each given flavor) are produced with
equal helicities. Hence nontrivial correlations of polarization
characteristics in the production of strange particles are to
be expected.\\
Manifestations of the 't~Hooft mechanism, due to its universality,
could be observed in a wide variety of processes involving
strangeness production in interactions
of hadrons (nucleons and $\pi$-mesons) with nucleons. However,
corresponding reactions occur at higher energies and thus the
qualitative signals induced by the presence of the 't~Hooft mechanism
might be distorted. Detailed analysis of experimental data on the
basis of the instanton induced mechanism will be published elsewhere.

\section{Acknowledgments}
Correspondence with W.Eyrich and Yu.Kalashnikova  is gratefully
acknowledged. We would like to thank the Institute for Theoretical
Physics~II at Ruhr--University Bochum for kind hospitality.

\end{document}